# Data-Calibrated Climate Outage Stress Testing for Joint Power-Communication Networks

Yoneke Graham
*Department of Computer Science*
*University of Nevada, Las Vegas*
Las Vegas, USA
grahay2@unlv.nevada.edu

Gelila Webster
*Department of Computer Science*
*University of Nevada, Las Vegas*
Las Vegas, USA
gelila.webster@unlv.edu

Tina Tran
*Department of Computer Science*
*University of Nevada, Las Vegas*
Las Vegas, USA
trant42@unlv.nevada.edu

Sohini Roy
*Department of Computer Science*
*University of Nevada, Las Vegas*
Las Vegas, USA
sohini.roy@unlv.edu

***Abstract*—Climate-driven outages pose a growing threat to cyber-physical power system (CPPS) resilience, particularly as modern grids increasingly rely on communication, sensing, and control infrastructure for situational awareness and coordinated response. Empirical outage studies and interdependency-aware resilience analyses are often studied separately, limiting our understanding of how observed climate-risk patterns translate into cascading degradation across coupled cyber and physical layers. This paper presents a data-calibrated stress-testing framework for joint power-communication networks. Using EAGLE-I outage records from 2015–2023, we characterize national climate-related outage patterns and identify severe-risk contexts defined by outage duration and customer impact and associated with event type, season, and geography. EAGLE-I customer-impact percentiles are log-normalized and scaled into representative stress intensities for spatially clustered benchmark scenarios on an IEEE 118-bus power system with an overlaid communication network. The Modified Implicative Interdependency Model (MIIM) is used to simulate cross-layer cascade propagation and quantify post-event operability, resilience gaps, and affected cyber-physical entities. Results show that even the highest-severity benchmark scenario degrades post-cascade operability to approximately 60%, producing a resilience gap roughly twice that of the inland baseline. The findings suggest that resilience assessment based only on outage statistics may miss bounded but meaningful amplification effects in interdependent power-communication infrastructure. The proposed framework provides an initial step toward data-calibrated, interdependency-aware resilience assessment for extreme-weather-affected CPPS.**

***Keywords*—*Cyber-physical power systems, climate-driven outages, data-calibrated stress testing, interdependent power-communication networks, cascading failures, grid resilience.***

## I. INTRODUCTION

The U.S. power grid faces growing disruption from extreme weather and natural disasters. Prior work identifies extreme events as a critical challenge for power-system planners and operators, motivating realistic resilience assessment under severe operating conditions [1]. Nationwide empirical evidence further indicates that U.S. outages are becoming more frequent, prolonged, and geographically uneven [2]. These trends make climate-related outages an increasingly important concern for grid reliability, resilience, and cyber-physical power system (CPPS) continuity.

This risk cannot be understood from the power layer alone. Modern power systems depend on communication, sensing, and control infrastructure for monitoring, situational awareness, coordination, and system response. A geographically concentrated climate event may therefore disrupt both power-system components and co-located communication entities that support grid observability and control. However, empirical outage characterization and interdependency-aware cascade simulation are generally investigated separately. Consequently, observed climate-outage patterns are rarely translated into benchmark stress scenarios for evaluating failure propagation and operability loss across coupled power and communication infrastructure.

To address this gap, this paper presents a data-calibrated CPPS stress-testing framework that links EAGLE-I outage evidence from 2015–2023 with cascade analysis of an interdependent IEEE 118-bus power-communication network. Climate-related outage patterns are characterized across event type, time, geography, and customer impact. An interpretable severity-risk model identifies event contexts associated with elevated severe-outage risk, while log-normalized customer-impact percentiles are mapped to bounded stress fractions and applied as spatially clustered initial failures. Cross-layer propagation is evaluated using the Modified Implicative Interdependency Model (MIIM) [3], which represents structural and operational dependencies through interdependency relations and assigns each cyber-physical entity one of three states: fully operational, operating at a reduced level, or failed. The principal contributions are: 1) a unified framework connecting large-scale EAGLE-I climate-outage characterization with MIIM-based resilience analysis; 2) an interpretable severity-risk modeling and scenario-construction procedure incorporating event type, geographic context, severity level, and customer-impact percentiles; and 3) an IEEE 118-bus benchmark evaluation quantifying post-cascade operability, resilience gaps, affected entities, component-level failure patterns, and cross-layer propagation under four representative climate-risk scenarios. MIIM itself is not introduced as a new model; the contribution lies in integrating it with empirically calibrated climate-risk scenarios. The IEEE 118-bus system is used as a controlled benchmark and is not intended to reproduce the physical geography of the coastal and inland groups identified in the EAGLE-I analysis.

## II. RELATED WORK

Prior research relevant to this study falls into three related strands: climate-driven outage analysis, data-driven outage-risk modeling, and interdependent cyber-physical infrastructure resilience analysis. Recent studies have established that weather- and climate-related disruptions are a growing concern for power-system reliability and resilience. Simulation-oriented work has examined distribution-grid response under extreme temperatures and weather conditions, emphasizing the need for realistic stress modeling under severe operating environments [1]. Nationwide outage studies have characterized large-scale temporal and spatial outage patterns across the United States [2]. Broader resilience-planning reviews further emphasize that extreme weather is a major driver of distribution-system disruption and that resilience analysis should consider interactions among critical infrastructure components [4].

A second stream of work has used statistical and machine-learning models for outage prediction and risk assessment. Prior studies have applied ensemble learning, Gaussian processes, support vector machines, and related methods to predict line outages during extreme meteorological events [5]. Other work has developed deep-learning models for weather-related outage prediction using weather, socio-economic, demographic, and infrastructure data [6]. Machine-learning approaches have also been used to predict outage duration for outage management and restoration planning [7]. These studies demonstrate the value of data-driven outage modeling, but their primary objective is usually predictive performance.

A third body of work examines cascading failures in interdependent infrastructures, particularly coupled power and communication systems. Prior studies have shown that interdependent communication and power distribution networks can be vulnerable to cascading failures and that communication-network structure can strongly influence robustness [8]. The Modified Implicative Interdependency Model (MIIM) represents intra- and interdependencies between power and communication entities using multi-state operability, enabling analysis of degraded as well as failed system states [3]. MIIM-based analysis has also been applied to smart-grid cyber-physical applications such as false-data injection and state-estimation resilience [9].

Taken together, prior work has advanced empirical outage characterization, outage-risk prediction, and interdependency-aware cascade modeling, but these strands are rarely linked in a single data-calibrated CPPS stress-testing pipeline. This paper connects them by linking EAGLE-I outage evidence, interpretable severity-risk modeling, and MIIM-based cascade simulation on a benchmark joint power-communication network.

## III. Dataset and Empirical Outage Characterization

### A. Dataset and Preprocessing

This study uses annual EAGLE-I (Environment for Analysis of Geo-Located Energy Information) outage files maintained by Oak Ridge National Laboratory (ORNL) [10]. The files were concatenated, available date fields were converted to datetime format, and year and month were derived from the reported outage-start time. Records from 2014 were excluded to avoid partial-year bias, leaving 525,360 records from 2015–2023. Unparseable dates were recorded as missing; no additional event-level deduplication or general missing-value imputation was performed in the downstream analyses.

Raw event types were consolidated into standardized categories. Climate-related outages are defined as Severe Weather, Winter Storm, or Natural Disaster. The main variables are event category, year, state, outage duration, and peak customers affected. For the temporal analyses of duration and customer impact, observations above the corresponding 95th percentile were excluded to limit upper-tail influence. Outage severity in this section is evaluated primarily through peak customers affected; the binary severe/non-severe label is defined in Section IV.

States were grouped as coastal or inland, with the coastal group comprising 30 states bordering an ocean, the Gulf of Mexico, or the Great Lakes, including Alaska and Hawaii. Frequency comparisons use outage counts aggregated by state, whereas customer-impact comparisons use the corresponding outage records. This grouping does not geographically map to the IEEE 118-bus benchmark. Because related records may share a state, year, or broader disturbance, the statistical findings are interpreted as dataset-level associations rather than causal or population-normalized estimates.

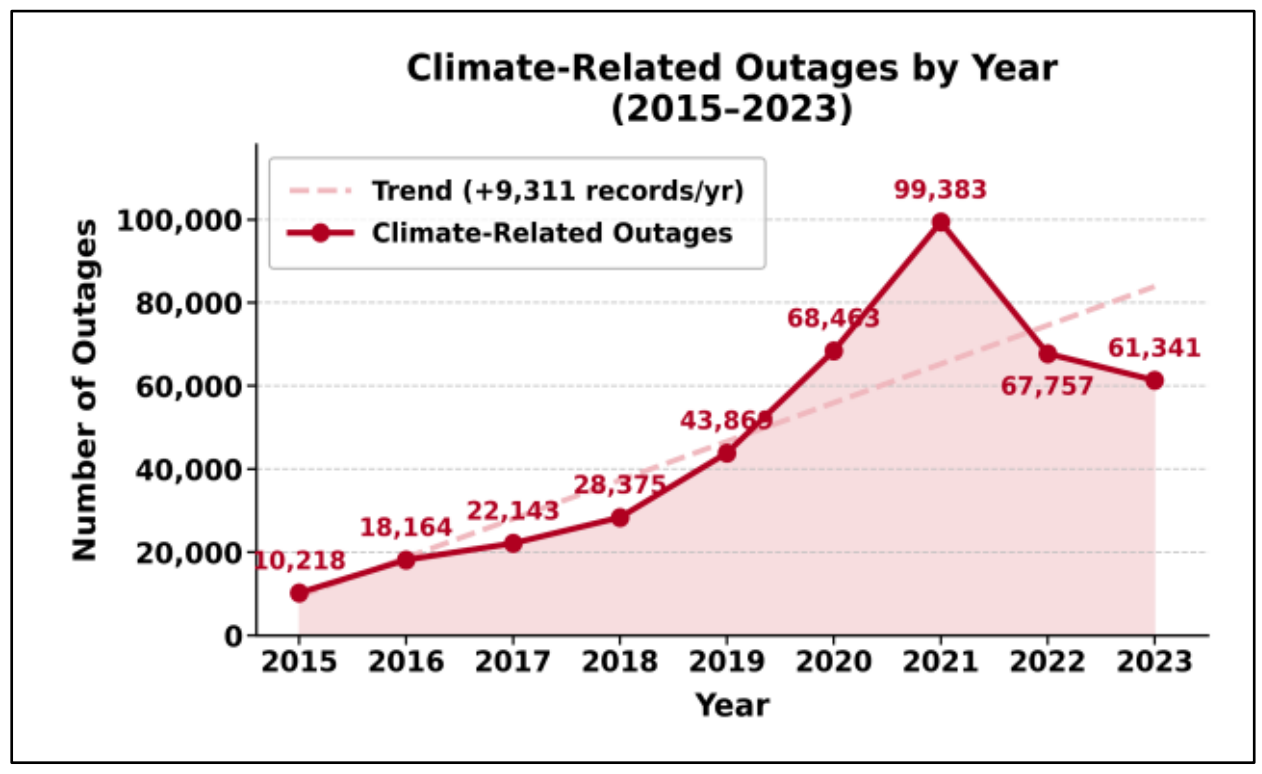

Fig. 1. Annual climate-related outage counts in the EAGLE-I dataset (2015–2023), showing a strong upward trend with a peak in 2021.

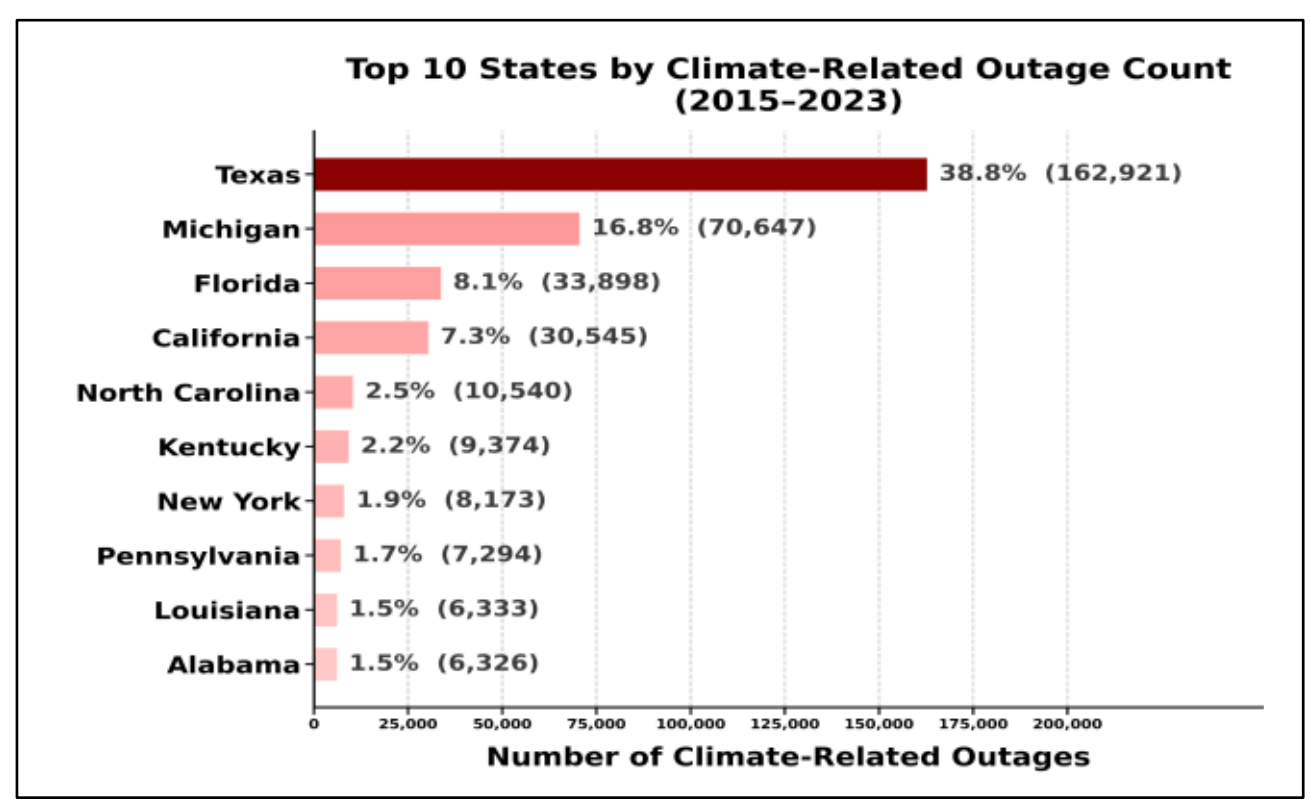

Fig. 2. Top 10 states by climate-related outage count (2015–2023)

### B. Descriptive Characterization

Climate-related outages account for 79.9% of all records in the cleaned dataset, with Severe Weather substantially exceeding Natural Disaster and Winter Storm in total count. As shown in Fig. 1, annual climate-related records increase from 10,218 in 2015 to a peak of 99,383 in 2021, followed by a decline in 2022 and 2023 while remaining above early-period levels. The fitted trend corresponds to an average increase of approximately 9,300 records per year over 2015–2023, indicating a positive temporal association within the analyzed dataset.

Climate-related records are also geographically concentrated. As shown in Fig. 2, Texas, Michigan, Florida, California, and North Carolina contribute a large share of the records, indicating substantial variation in recorded outage frequency across states. Because the counts are not normalized by state population, customer base, or reporting coverage, they should not be interpreted as direct measures of relative vulnerability. The customer-impact distribution is strongly right-skewed: most records involve relatively few customers, while a smaller upper tail accounts for disproportionately large impacts, particularly for Severe Weather. Coastal outage records also exhibit significantly greater mean customer impact than inland records. These findings motivate the use of event type, geographic context, and customer-impact percentiles in constructing the EAGLE-I-calibrated benchmark stress scenarios.

### C. Hypothesis Testing

To move beyond descriptive patterns, six exploratory hypotheses were evaluated using OLS regression and two-sample t-tests, as summarized in Table I. H1(a) shows a significant positive trend in annual climate-related outage frequency. H1(b) shows no significant trend in the annual proportion of major outages. H1(c) and H1(d) produce statistically significant coefficients after excluding observations above the corresponding 95th percentiles; however, their near-zero $R^2$ values indicate negligible explanatory power. Thus, outage duration shows only a weak positive temporal association, while customer impact shows a weak negative association. H2(a) shows higher mean outage frequency among coastal states, but the difference is not statistically significant. H2(b) shows significantly greater mean customer impact for coastal than inland outage records ($t = 14.88, p < 0.001$).

Overall, the analysis identifies three empirical patterns used in scenario design: increasing climate-related outage frequency, the dominance of Severe Weather in record count and upper-tail customer impact, and higher customer impact among coastal outage records. These findings motivate the use of event type, geographic context, and customer-impact percentiles in constructing EAGLE-I-calibrated benchmark scenarios for interdependency-aware CPPS cascade simulation. Because the tests do not fully account for temporal and geographic dependence, the results are interpreted as exploratory dataset-level associations.

TABLE I. HYPOTHESIS TESTING RESULTS – EAGLE-I DATASET, 2015-2023 (ALPHA = 0.05)

| Hypothesis | Test | Key Statistic | Outcome |
|---|---|---|---|
| H1(a): Frequency increased over time | OLS | $\beta = 9311$ records/yr, $R^2 = 0.78$, $p < 0.05$ | Significant positive trend |
| H1(b): Major-outage proportion remained stable | OLS | $\beta = -0.0018$, $R^2 = 0.025$, $p > 0.05$ | No significant trend |
| H1(c): Average duration shows weak trend | OLS | $\beta = 0.016$, $R^2 \approx 0, p < 0.05$ | Significant but negligible positive trend |
| H1(d): Average customer-impact severity shows weak trend | OLS | $\beta = -26.91$, $R^2 \approx 0, p < 0.05$ | Significant but negligible negative trend |
| H2(a): Coastal frequency > inland | Two-sample t-test | $t = 1.66, p = 0.10$ | Not statistically significant |
| H2(b): Coastal customer-impact severity > inland | Two-sample t-test | $t = 14.88, p < 0.001$ | Significantly higher for coastal records |

## IV. SEVERITY RISK MODELING

### A. Target Definition and Feature Set

To identify contexts associated with elevated severe-outage risk, a binary severity label was constructed from the cleaned EAGLE-I dataset. Unlike Section III, which evaluates customer impact primarily through peak customers affected, the severity-risk model combines outage duration and customer impact. Each variable was independently normalized to $[0, 1]$, and their average defined a composite severity score. Records in the top quartile of this score were labeled Severe (1), while the remaining records were labeled Non-Severe (0), yielding 131,340 severe records (25%) and 394,020 non-severe records (75%). The 75th-percentile threshold was selected as a transparent upper-tail screening criterion that retains sufficient severe observations for stable estimation; it is not intended as a regulatory or utility-defined major-outage threshold.

Predictors include event category, year, season, region, coastal status, and selected season-region interaction terms. Duration, peak customers affected, and the composite severity score were excluded from the feature set to prevent target leakage. The resulting features were encoded and standardized before model fitting to support interpretable comparison of severe-risk associations across outage contexts.

### B. Logistic Regression Setup

The labeled dataset was divided using an 80/20 stratified train–test split to preserve the severe/non-severe class ratio. A logistic regression model with L2 regularization and balanced class weighting was fitted to the training set to account for the 75/25 class imbalance. For outage record $i$, the probability of severe impact is modeled as:

$$P(Y_i = 1 \mid x_i) = \frac{1}{1 + \exp[-(\beta_0 + \beta^T x_i)]} \quad (1)$$

where $Y_i = 1$ denotes a severe record and $x_i$ contains the contextual predictors defined in Section IV-A. Logistic regression was selected because its signed coefficients provide an interpretable ranking of event types, seasons, and geographic contexts associated with elevated severe-outage risk. These coefficient-level associations inform the event contexts represented in Section V, while scenario footprint sizes are calibrated separately using EAGLE-I customer-impact percentiles.

Performance was evaluated on the held-out test set using accuracy, precision, recall, and AUC-ROC. This evaluation measures record-level discrimination under the random stratified split and does not test transfer to unseen years or independently grouped disturbance events. The ranked coefficient estimates are shown in Fig. 3.

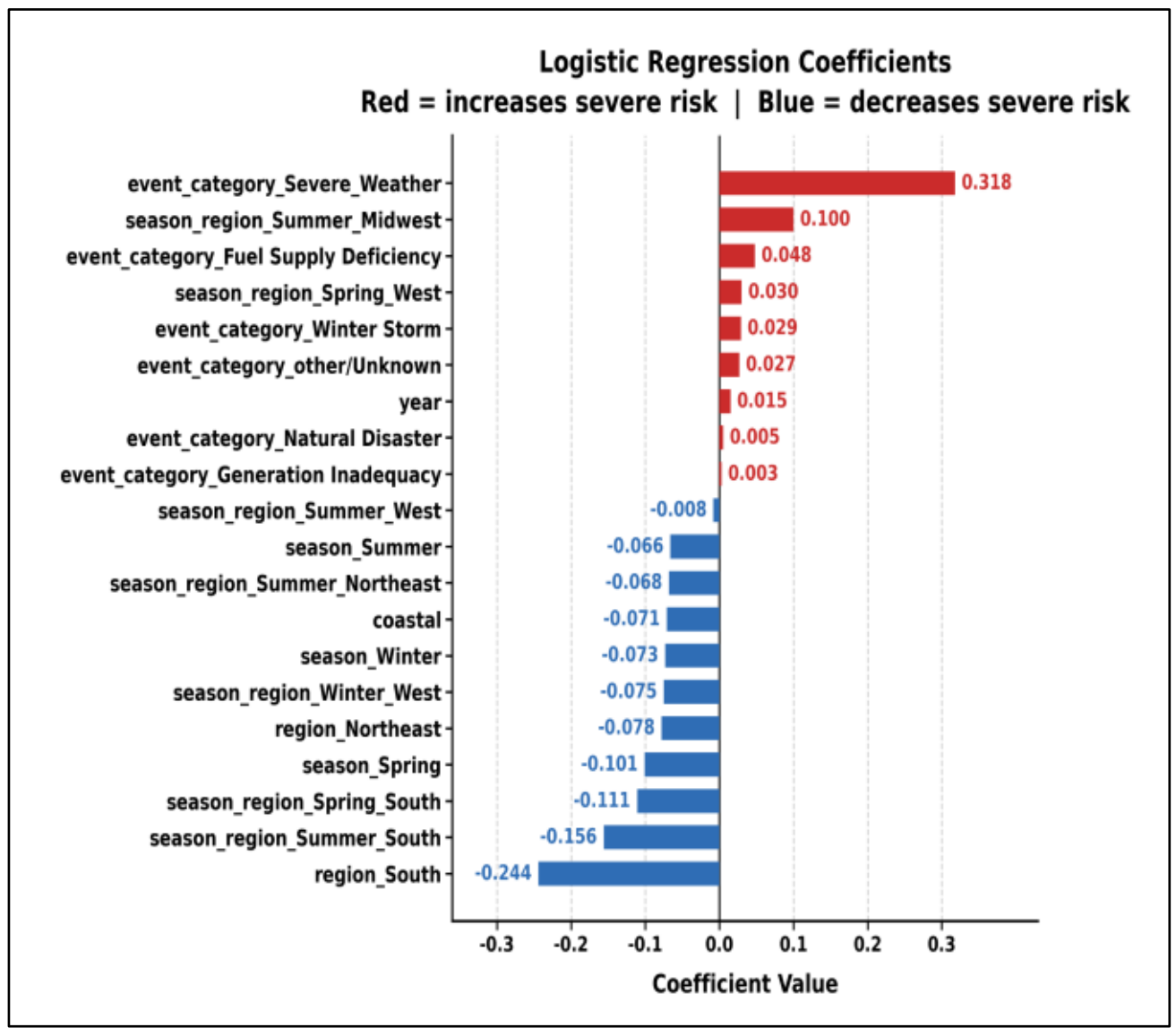

Fig. 3. Logistic regression coefficient plot showing the strongest positive and negative predictors of severe outage risk.

## C. Results and Interpretation

On the held-out test set of 105,072 records, the model achieves accuracy = 0.664, severe-class precision = 0.363, severe-class recall = 0.454, and AUC-ROC = 0.634. This modest discrimination is consistent with the model's intended role as an interpretable context-ranking layer rather than an operational outage predictor. Duration and peak customers affected were used to construct the severity label and were therefore excluded from the predictors to prevent target leakage, which also limits predictive performance.

TABLE II. LOGISTIC REGRESSION CLASSIFICATION REPORT

| Class | Precision | Recall | F1-score | Support |
|---|---|---|---|---|
| Non-Severe | 0.80 | 0.73 | 0.77 | 78,804 |
| Severe | 0.36 | 0.45 | 0.40 | 26,268 |
| Accuracy | – | – | 0.66 | 105,072 |
| Macro-average | 0.58 | 0.59 | 0.58 | 105,072 |

As shown in Fig. 3, Severe Weather has the largest positive coefficient, followed by smaller positive coefficients for Winter Storm and Natural Disaster. The coastal indicator is negative after adjustment, although Section III shows higher mean customer impact among coastal outage records. This reflects a conditional association after accounting for event category, season, and region and does not invalidate the simpler coastal–inland comparison.

Overall, the model provides adjusted, interpretable associations for prioritizing representative event contexts. It does not physically calibrate the cascade scenarios; scenario footprint sizes are determined separately from EAGLE-I customer-impact percentiles in Section V.

# V. MIIM-BASED SCENARIO CONSTRUCTION AND SIMULATION SETUP

## A. Rationale for MIIM-Based Interdependency Analysis

The Modified Implicative Interdependency Model (MIIM) [3] represents a joint power-communication network as $J(E, F(E))$, where $E = P \cup C \cup CP$ contains power, communication, and cross-layer entities, and $F(E)$ is the set of Inter-Dependency Relations (IDRs) describing their operational dependencies. Each entity $i$ has a state $s_i \in \{0, 1, 2\}$, denoting failed, degraded, or fully operational status, respectively. Following scenario initialization, entity states are updated iteratively as:

$$s_i^{(t+1)} = F_i\left(s^{(t)}\right) \tag{2}$$

until no further state changes occur. MIIM uses multi-valued logical operators to represent conjunctive, alternative, and partial dependencies, enabling cross-layer cascade analysis without treating every disruption as a binary failure. The complete operator definitions and comparison with the binary Implicative Interdependency Model are provided in [3]. In this work, MIIM is used as the simulation engine for translating EAGLE-I-calibrated climate-risk scenarios into operability loss in an IEEE 118-bus joint power-communication network; the contribution is data-calibrated stress testing rather than a new interdependency model.

## B. Scenario Construction Logic

The scenarios are representative climate-risk contexts rather than reconstructions of specific historical events. Their event types and geographic labels are informed by Sections III and IV: Severe Weather dominates the outage records and has the largest positive severity-risk coefficient, Winter Storm also shows a positive association, and coastal outage records exhibit higher mean customer impact than inland records. The scenarios therefore span dominant event types, coastal and inland contexts, and moderate and extreme customer-impact levels while remaining suitable for controlled benchmark analysis.

Scenario magnitude is calibrated using EAGLE-I customer-impact percentiles. Because customer impact is strongly right-skewed, the selected percentile value $C$ is log-normalized as:

$$t = {(\ln C - \ln C_{min})}\Big/{(\ln C_{max} - \ln C_{min})} \tag{3}$$

where $C_{\min}$ and $C_{\max}$ are the minimum and maximum selected percentile values across the scenario set. The normalized value is mapped to the affected-network fraction:

$$f = 0.15 + 0.20t \tag{4}$$

and rounded to the nearest 5% points. Applied to the 107 modeled substations in the IEEE 118-bus joint network, the resulting fractions of 0.15, 0.20, 0.25, and 0.35 correspond to 16, 21, 27, and 37 initially affected substations, respectively. The mapping is a monotonic, bounded benchmark-scaling rule: log normalization preserves the empirical severity ordering while limiting domination by the largest customer-impact value. The 15%–35% range avoids both minimally disruptive footprints and near-system-wide initialization that would obscure cascade behavior. It is not a physical conversion from customers affected in EAGLE-I to failed IEEE 118-bus substations, and the coastal and inland labels do not assign literal geography to the benchmark network.

## C. Four Scenarios

Four representative scenarios are considered in the MIIM analysis. Their empirical calibration inputs and corresponding initialization parameters are summarized in Table III. The table reports the event type, geographic context, selected EAGLE-I customer-impact percentile, customer-impact value at that percentile, calibrated affected-network fraction, number of initially affected substations, and mean number of cyber-physical entities initialized as failed at $T_0$.

TABLE III. SCENARIO PARAMETERS: DERIVED FROM EAGLE-I CUSTOMER-IMPACT PERCENTILE ANALYSIS

| Scenario | Event Type | Percentile | Customer Impact at Percentile | Stress fraction (f) | Substations Initially Affected | Initial Entities Affected (at T0) |
|---|---|---|---|---|---|---|
| S1: Moderate Coastal Severe Weather | Severe Weather | p75 | 1933 | 20% | 21 | 90 |
| S2: Extreme Coastal Severe Weather | Severe Weather | p90 | 5946 | 35% | 37 | 158 |
| S3: Inland Natural Disaster | Natural Disaster | p75 | 1451 | 15% | 16 | 68 |
| S4: Coastal Winter Storm | Winter Storm | p90 | 3170 | 25% | 27 | 111 |

*Initial entities affected at T0 are reported as means over three spatially clustered footprints per scenario.

S1 and S2 provide a moderate-versus-extreme comparison for coastal Severe Weather, while S3 serves as the lower-stress inland comparison and S4 represents a high-percentile coastal Winter Storm. Scenarios at the same percentile may have different affected fractions because the calibration uses the corresponding customer-impact value $C$, rather than percentile rank alone. Together, the scenarios span p75 and p90 severity levels, coastal and inland contexts, and the three climate-event categories identified in the empirical analysis.

## D. Initial Failure Design

In this study, MIIM [3] is applied to a joint power-communication network constructed by overlaying a communication-layer design onto the IEEE 118-bus power network. The communication layer includes substation servers, gateways, PMUs, and transport multiplexers, including S-ADMs and OADMs. This overlay represents a simplified utility communication architecture for benchmark CPPS stress testing rather than a utility-specific deployment. Gateways connect substations to the communication transport layer through Ethernet and DWDM channels, while the transport layer is dual-homed to two control centers through redundant communication paths.

Each scenario's initial failure set represents spatially clustered area damage rather than selective disruption of the power layer alone. For a given scenario, a spatially clustered set of buses is designated as directly damaged, with the number of affected substations determined by the calibrated affected-network fraction in Table III. The coastal and inland labels describe the empirical EAGLE-I contexts used for scenario calibration and do not assign literal geography to the IEEE 118-bus network. For each affected substation, all associated buses, the substation server, the co-located gateway, and any co-located PMUs are included in the initial failure set and assigned MIIM state 0.

Communication transport entities are initialized using a connectivity-based proxy for local co-location. An Ethernet or optical transport multiplexer, including an S-ADM or OADM, is included in the initial failure set when it has an Ethernet or DWDM channel to a gateway at a damaged substation. Connections to the two control-center gateways are excluded from this rule because they represent backbone homing paths rather than local substation co-location. This prevents a control-center uplink from causing a transport entity to be classified as locally damaged solely because of logical connectivity. Under this initialization rule, a transport multiplexer is directly failed only when its associated local substation lies within the damaged footprint.

This initialization captures the assumption that climate-driven disturbances can simultaneously damage co-located power and communication infrastructure while preserving communication redundancy outside the directly affected area. The IEEE 118-bus system serves only as the benchmark topology substrate and does not geographically correspond to the EAGLE-I records. To reduce sensitivity to a single selected footprint, each scenario is evaluated using three distinct spatially clustered footprints with the same calibrated affected-network fraction. Reported $T_0$ entity counts, operability values, resilience gaps, and total affected-entity counts are means across these three footprints.

## E. Evaluation Metrics

Six metrics are used to characterize cascade behavior. Initial Failed Entities counts the entities assigned state 0 at $T_0$, before cascade propagation. Cascade Depth is the number of synchronous MIIM update steps after $T_0$ required for the system state to reach steady state, such that no entity changes state in the subsequent update.

Post-cascade Operability, denoted by $O$, measures the mean normalized steady-state operability across the $N = 446$ modeled non-cable entities:

$$O = \left(1/_N\right) \sum \left({s_i}^*/_2\right) \tag{5}$$

where $s_i^* \in \{0, 1, 2\}$ is the steady-state condition of entity $i$, corresponding to failed, degraded, and fully operational states, respectively. Thus, $O \in [0,1]$, with larger values indicating greater retained system functionality. The Resilience Gap is defined as:

$$G = 1 - O \tag{6}$$

where a larger value indicates greater system-level degradation.

Total Affected Entities counts the entities that either directly failed at $T_0$ or reach a degraded or failed state by steady state. Finally, Post-cascade Failed Entities counts only entities that are fully failed at steady state, $s_i^* = 0$. This metric is reported in Fig. 7 and is a subset of Total Affected Entities because it excludes entities remaining in the degraded state.

## VI. MIIM-Based Resilience Analysis Results and Discussion

### A. Initial Damage Structure

The initial failure sets reflect the calibrated affected-network fraction and spatial footprint assigned to each representative climate-risk scenario. All $T_0$ entity counts are reported as means across three spatially clustered footprints per scenario. As expected from the initialization design, S2, the Extreme Coastal Severe Weather scenario, has the largest mean initial cross-layer failure set, with 158 entities initialized as failed across 37 substations. S3, the Inland Natural Disaster scenario, has the smallest, with 68 initially failed entities across 16 substations. S1 and S4 fall between these cases, with mean initial failure sets of 90 and 111 entities across 21 and 27 substations, respectively. All four scenarios affect both the power and communication layers at $T_0$ because each damaged substation includes its associated buses, substation server, co-located gateway, and any co-located PMUs. Transport multiplexers are included only when they are locally sited at damaged substations, as described in Section V-D. These differences represent the prescribed initial damage scale, while cascade-induced degradation and failure are evaluated separately in the subsequent analyses.

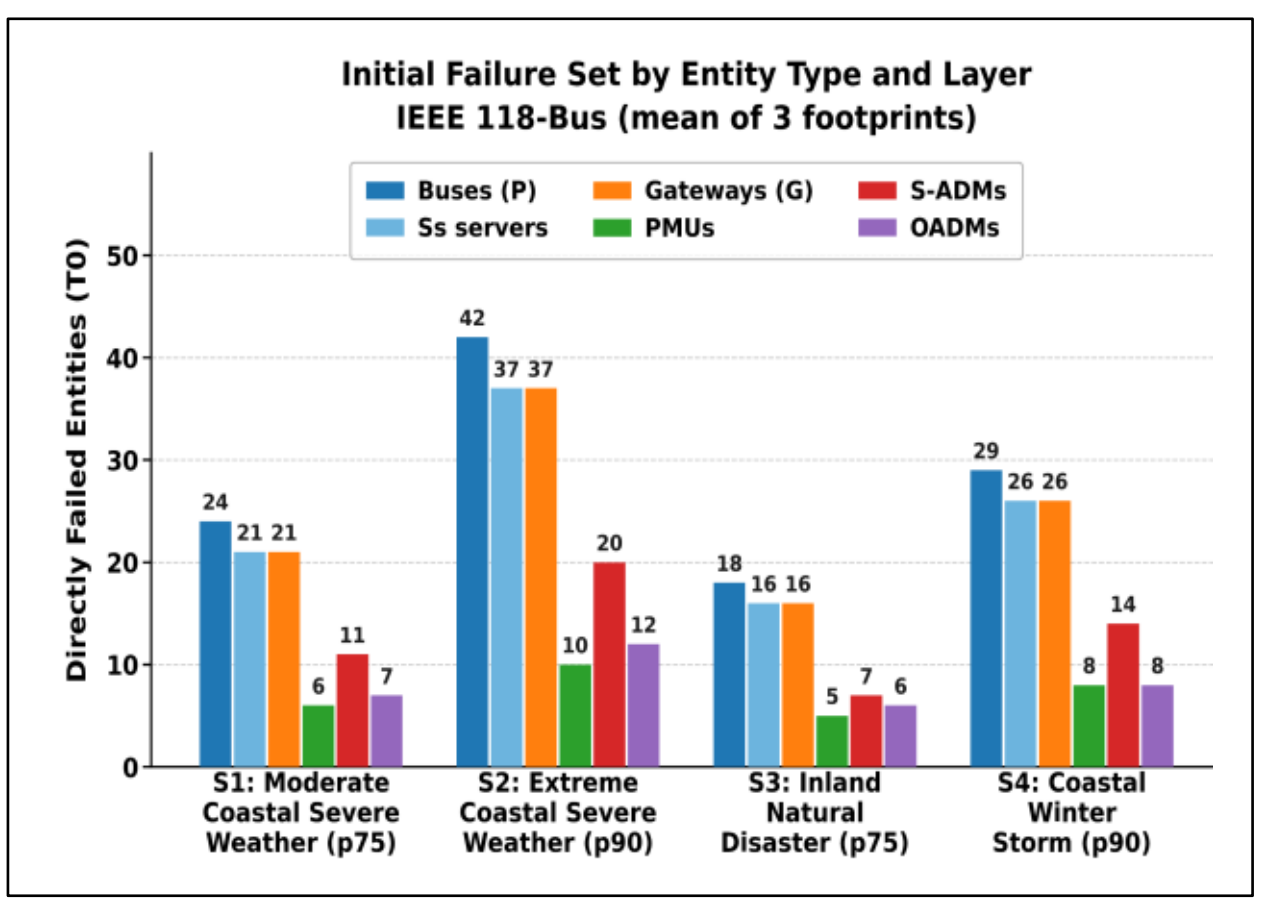

Fig. 4. Initial failure set breakdown by entity type across all four scenarios.

### B. System Level Operability

The overall resilience outcomes are summarized in Table IV and visualized in Fig. 5. All four calibrated scenarios reduce post-cascade operability relative to the fully operational baseline, with S2, the Extreme Coastal Severe Weather scenario, producing the greatest degradation, with operability $= 0.604$ and resilience gap $= 0.396$. S3, the Inland Natural Disaster scenario, produces the least degradation, with operability $= 0.816$ and resilience gap $=$ 0.184. S1 and S4 exhibit intermediate impacts, with operabilities of 0.768 and 0.722 and resilience gaps of 0.232 and 0.278, respectively. The resilience gap of S2 is approximately 2.1 times that of S3.

The MIIM simulations retain the ordering of the calibrated stress scenarios, $S2 > S4 > S1 > S3$, after cascade propagation. Scenarios initialized using higher EAGLE-I-calibrated customer-impact levels therefore produce greater post-cascade operability loss in the coupled power-communication benchmark. This consistency provides a systematic connection between the empirical scenario calibration and the resulting cyber-physical resilience assessment.

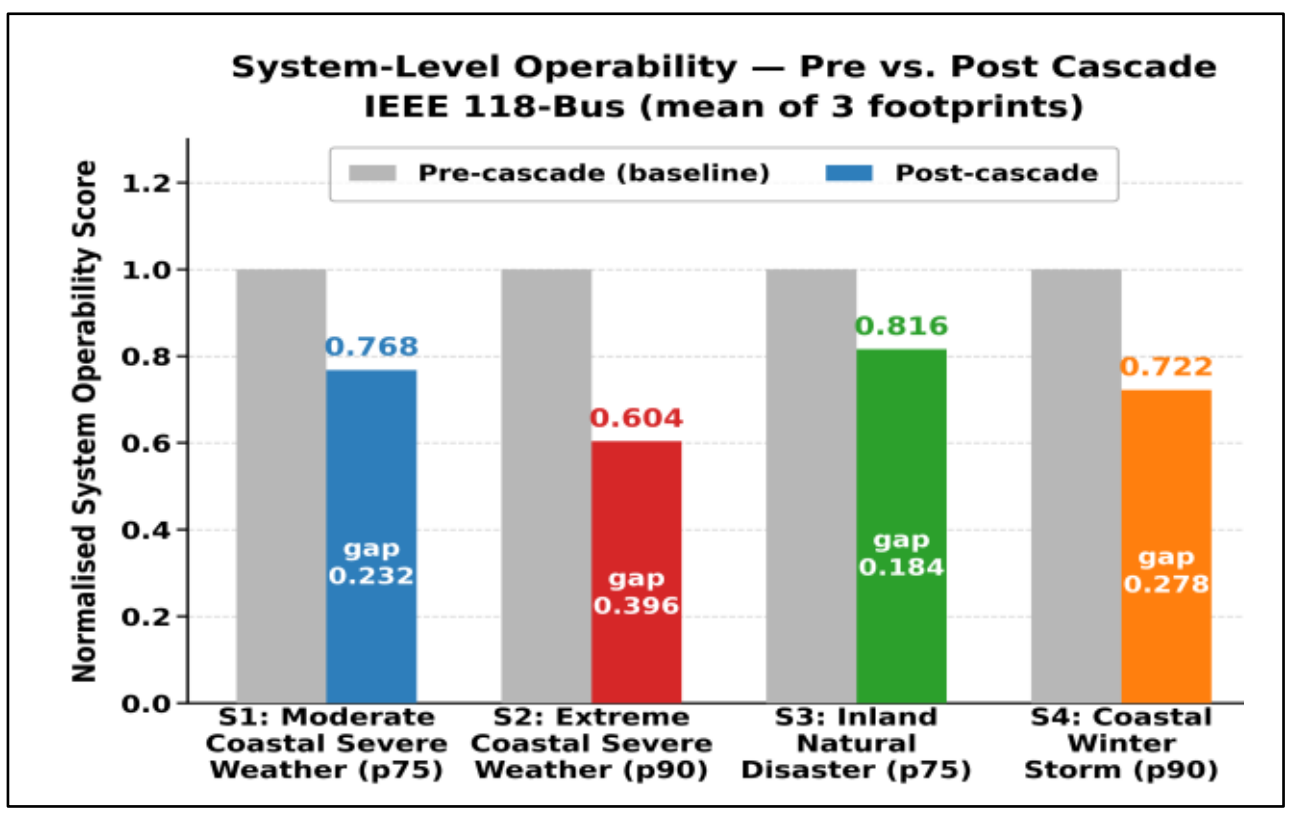

Fig. 5. System operability before and after cascade for all four scenarios, with resilience gaps indicated.

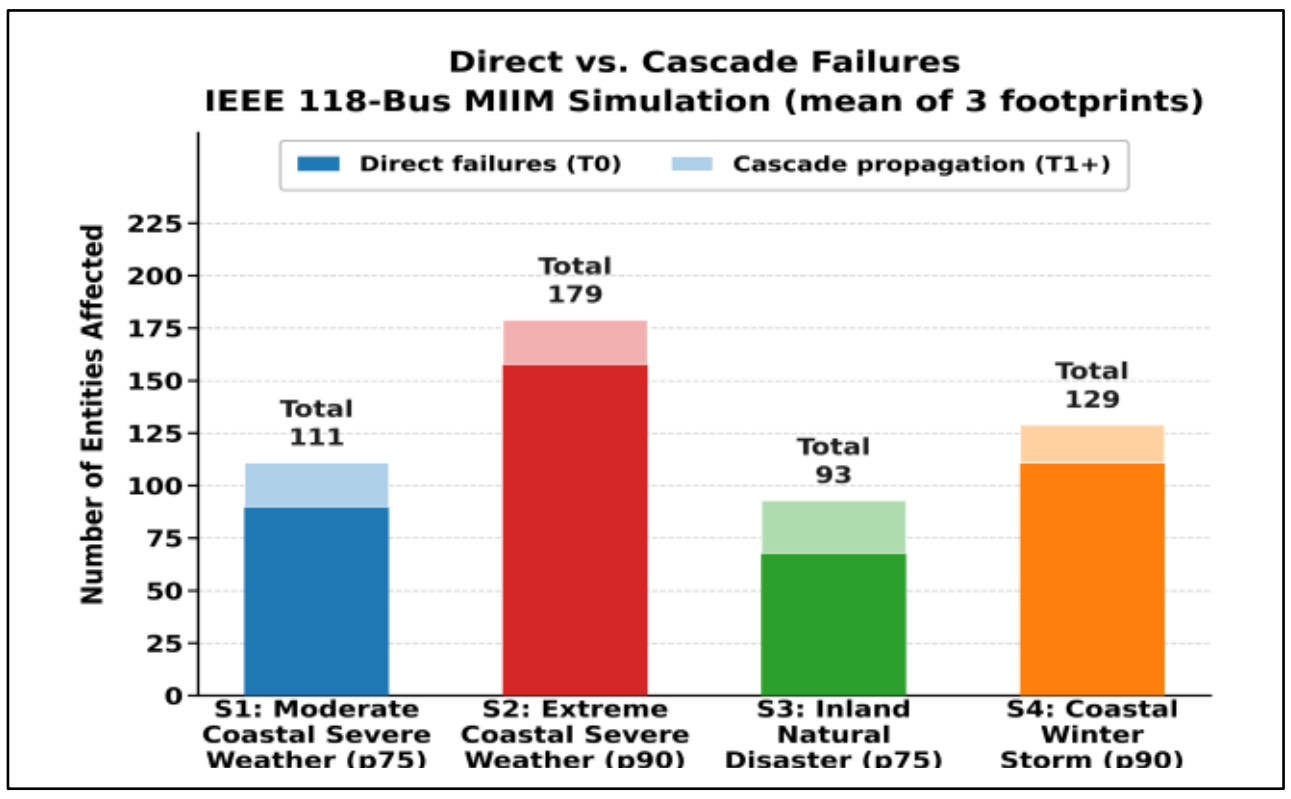

Fig. 6. Direct vs. cascade-propagated failures for each scenario

### C. Cascade Propagation Behavior

Fig. 6 decomposes total affected entities into entities directly failed at $T_0$ and additional entities affected through cascade propagation. Across all four scenarios, the cascade extends disruption beyond the initial spatial footprint but remains bounded, with each simulation reaching steady state within three propagation steps. S2 increases from 158 to 179 affected entities, while S1, S4, and S3 increase from 90 to 111, 111 to 129, and 68 to 93 entities, respectively. These changes correspond to 18–25 additional affected entities through cascade propagation. This bounded amplification indicates that, under the modeled topology, dependency relations, and redundancy assumptions, the coupled power-communication network contains the propagation of spatially clustered initial damage rather than producing runaway cascades. Because the cascade increment is similar across scenarios, differences in overall resilience are driven primarily by the calibrated initial damage footprint, which is based on EAGLE-I customer-impact percentiles, rather than by differences in propagation depth. The common three-step convergence should therefore be interpreted as a property of this benchmark configuration rather than a universal cascade-depth result. Accordingly, the findings demonstrate how data-calibrated disruption severity translates into interdependency-aware resilience degradation, rather than attributing the observed differences solely to event type or geographic context. Notably, S3 exhibits the largest absolute cascade increment despite being the least damaging scenario overall. This further illustrates that cascade amplification and total scenario severity are not identical: the calibrated initial damage scale primarily governs overall resilience loss, while cross-layer propagation provides a bounded additional effect.

TABLE IV. MIIM CASCADE SIMULATION RESULTS USING IEEE 118-BUS JOINT POWER-COMMUNICATION NETWORK

| Scenario | T0 Failed | Depth | Operability (O) | Resilience Gap (G) | Total Affected Entities |
|---|---|---|---|---|---|
| S1: Moderate Coastal Severe Weather | 90 | 3 | 0.768 | 0.232 | 111 |
| S2: Extreme Coastal Severe Weather | 158 | 3 | 0.604 | 0.396 | 179 |
| S3: Inland Natural Disaster | 68 | 3 | 0.816 | 0.184 | 93 |
| S4: Coastal Winter Storm | 111 | 3 | 0.722 | 0.278 | 129 |

### D. Component-Level Failure Patterns

Fig. 7 reports post-cascade fully failed entities, defined as state 0 at steady state, by entity type across all four scenarios. Substation server and gateway failures remain exactly matched because both are initialized as failed at directly damaged substations and each gateway depends on its co-located server, illustrating modeled power-communication coupling. S-ADM and OADM failures also increase as more substations are directly damaged, indicating greater exposure of locally associated communication equipment. These patterns support the bounded-propagation interpretation in Section VI-C: communication-layer assets fail alongside power-layer assets, but their failure counts generally scale with the damaged spatial footprint rather than indicating uncontrolled communication-backbone collapse.

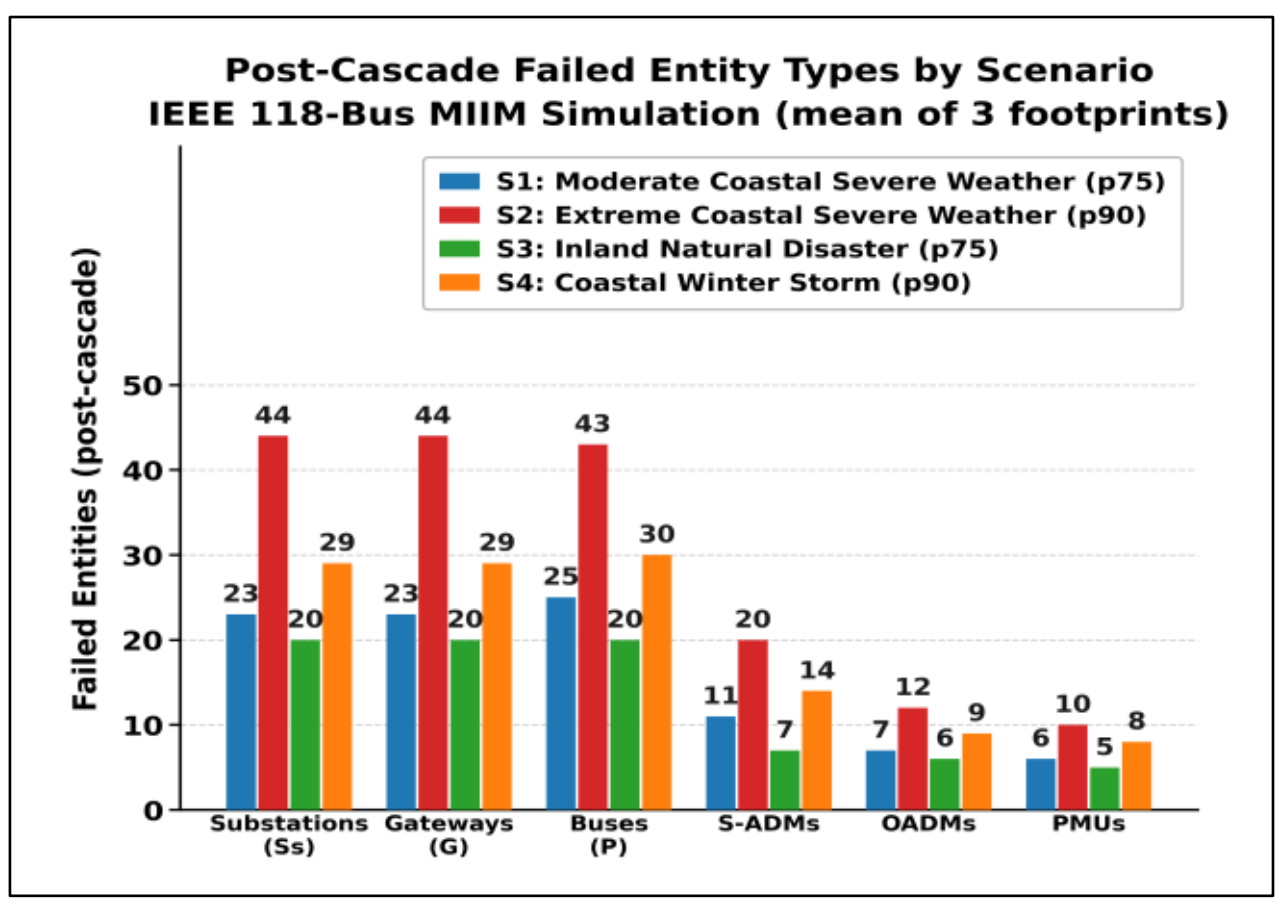

Fig. 7. Post-cascade failed entity counts by type across all four scenarios.

### E. Integrated Interpretation

Sections III–VI form a data-calibrated CPPS stress-testing pipeline linking empirical outage patterns, interpretable severity-risk modeling, scenario calibration, and MIIM simulation. Scenarios with larger calibrated initial damage footprints produce greater post-cascade resilience gaps, with S2 producing the greatest degradation and S3 the least. Because the scenarios differ in event type, empirical context, customer-impact percentile, and affected-network fraction, the results do not isolate a causal coastal effect. Instead, the framework shows how empirically informed climate-risk contexts translate into interdependency-aware operability loss under the modeled bounded-cascade conditions.

## VII. CONCLUSION

This paper presented a data-calibrated framework linking EAGLE-I outage characterization, interpretable severity-risk modeling, and MIIM cascade simulation for joint power-communication resilience assessment. Analysis of 2015–2023 records identified an increase of approximately 9,300 climate-related outages per year, the dominance of Severe Weather, and greater mean customer impact in coastal than inland states. These findings informed four customer-impact-calibrated scenarios on an IEEE 118-bus joint network. The Extreme Coastal Severe Weather scenario reduced post-cascade operability to approximately 0.60 and produced a resilience gap roughly twice that of the Inland Natural Disaster scenario, while cross-layer cascade amplification remained bounded. The results support jointly considering power, communication, and control infrastructure in resilience planning, particularly for coastal risk contexts. Future work will examine geographically grounded utility-scale topologies, alternative initialization assumptions and interdependency models, broader spatial-footprint sensitivity, and MIIM-based mitigation strategies.

## REFERENCES

[1] S. Hanif, M. Mukherjee, S. Poudel, M. G. Yu, R. A. Jinsiwale, T. D. Hardy, and H. M. Reeve, “Analyzing at-scale distribution grid response to extreme temperatures,” *Applied Energy*, vol. 337, p. 120886, May 2023.

[2] B. Li, J. Ma, O. A. Omitaomu, and A. Mostafavi, “Revealing growing and disparate vulnerability in the U.S. power system: A spatiotemporal analysis of nationwide outages from 2014 to 2023,” *International Journal of Disaster Risk Reduction*, vol. 133, p. 105980, 2026.

[3] S. Roy, H. Chandrasekaran, A. Pal and A. Sen, "A New Model to Analyze Power and Communication System Intra-and-Inter Dependencies," *2020 IEEE Conference on Technologies for Sustainability (SusTech)*, Santa Ana, CA, USA, 2020, pp. 1-8.

[4] S. Paul, A. Poudyal, S. Poudel, A. Dubey, and Z. Wang, “Resilience assessment and planning in power distribution systems: Past and future considerations,” *Renewable and Sustainable Energy Reviews*, vol. 188, p. 113859, 2023.

[5] A. Unlu and M. Peña, “Assessment of Line Outage Prediction Using Ensemble Learning and Gaussian Processes During Extreme Meteorological Events,” *Wind*, vol. 4, no. 4, pp. 342–362, 2024.

[6] X. Wang, N. Fatehi, C. Wang, and M. H. Nazari, “Deep Learning-Based Weather-Related Power Outage Prediction with Socio-Economic and Power Infrastructure Data,” in *Proc. IEEE Power & Energy Society General Meeting (PESGM)*, 2024, pp. 1–5.

[7] B. Ghasemkhani, R. A. Kut, R. Yilmaz, D. Birant, Y. A. Arıkök, T. E. Güzelyol, and T. Kut, “Machine Learning Model Development to Predict Power Outage Duration (POD): A Case Study for Electric Utilities,” *Sensors*, vol. 24, no. 13, p. 4313, 2024.

[8] W. K. Chai, V. Kyritsis, K. V. Katsaros, and G. Pavlou, “Resilience of Interdependent Communication and Power Distribution Networks against Cascading Failures,” in *Proc. IFIP Networking Conference*, 2016, pp. 37–45.

[9] S. Roy and A. Sen, "Identification and Mitigation of False Data Injection using Multi State Implicative Interdependency Model (MSIIM) for Smart Grid," *2021 IEEE International Conference on Communications Workshops (ICC Workshops)*, Montreal, QC, Canada, 2021, pp. 1-6.

[10] V. Tansakul et al., “EAGLE-I Power Outage Data Information,” Oak Ridge National Laboratory, Oak Ridge, TN, USA, dataset, 2023, doi: 10.13139/ORNLNCCS/1975203.